\documentclass[
aps,pra,twocolumn,superscriptaddress,%
amsmath,amssymb
]%
{revtex4-1}

\usepackage{hyperref}
\usepackage{lipsum}
\usepackage{graphicx}
\usepackage{color,soul}
\usepackage{bm}
\usepackage{verbatim}

\graphicspath{{./}{./figs/}{./figs/work/}{./figs/log/}}
\usepackage{cancel}

\newcommand{\al}{{\itshape et al.} }
\newcommand{\ie}{{\itshape i.e. } }
\newcommand{\eg}{{\itshape e.g. } }
\begin{document}
\title{Phonon-mediated spin-spin interactions between trapped Rydberg atoms}
\author{R.~V.~\surname{Skannrup}} 
\email{r.u.skannrup@tue.nl}
\affiliation{Eindhoven University of Technology, P.O. Box 513, %
5600 MB  Eindhoven, The Netherlands}
\author{R.~\surname{Gerritsma}}
\affiliation{Van der Waals-Zeeman Institute of Physics, %
University of Amsterdam, 1098 XH Amsterdam, The Netherlands}
\author{S.~J.~J.~M.~F.~\surname{Kokkelmans}}
\affiliation{Eindhoven University of Technology, P.O. Box 513, %
5600 MB  Eindhoven, The Netherlands}
\date{\today}

\begin{abstract}
We theoretically investigate the possibility of creating phonon-mediated 
spin-spin interactions between neutral atoms trapped in optical tweezers. 
By laser coupling the atoms to Rydberg states, collective 
modes of motion appear. We show that these can be used to mediate 
effective spin-spin interactions or quantum logic gates between the 
atoms in analogy to schemes employed in trapped ions. In particular, 
we employ Rydberg dressing in a novel scheme to induce the needed 
interaction, and we show that it is possible to replicate the working 
of the M\o{}lmer-S\o{}rensen entanglement scheme. The 
M\o{}lmer-S\o{}rensen gate is widely used in emerging quantum 
computers using trapped ion qubits and currently features some of the 
highest fidelities of any quantum gate under consideration. 
We find arbitrarily high fidelity for the coherent time evolution 
of the two-atom state even at non-zero temperature.
\end{abstract}
\pacs{}
\maketitle


The quest for scalable, high fidelity quantum logic gates 
is on \cite{flagship}. 
State-of-the-art quantum gates based on trapped ions show the 
best fidelities in the field of quantum logic.
A notable quantum gate protocol, inspiring 
this work, is the so-called M\o{}lmer-S\o{}rensen (MS) gate
\cite{PhysRevLett.82.1971,PhysRevLett.86.3907,PhysRevA.62.022311}, 
which uses trapped ions to create a quantum gate. It is based on 
phonon-mediated interactions, and in combination with the Hadamard- 
and $\pi/2$-gates the MS gate can be used to implement a 
C-NOT gate.
This gate has been experimentally realized 
and has shown very high fidelities 
\cite{Sackett2000,Leibfried2003,Benhelm2008,Kirchmair_2009,%
PhysRevLett.117.060505,PhysRevLett.117.060504},
but trapped ion gates lack in terms of scalability, as it is difficult 
to control many trapped ions. 
On the other hand, quantum gates using neutral, highly excited 
Rydberg atoms 
\cite{PhysRevLett.85.2208,PhysRevA.92.022336,PhysRevLett.104.010503,%
PhysRevLett.87.037901,PhysRevA.82.030306,PhysRevLett.104.010502,%
PhysRevLett.74.4091,Jau2015}, 
constitute a much more scalable platform
\cite{Saffman2016,RevModPhys.82.2313}, but show significantly 
lower experimental fidelities.
Rydberg atom quantum gates rely on strong dipole-dipole 
interactions between electrically neautral Rydberg atoms to 
facilitate entanglement. 

These considerations raise the question: "Can phonon-mediated
interactions be used to implement quantum gates between neutral atoms 
in a similar way as between ions?" In
this paper we will justify that the answer is "yes" and
we present both a model and a recipe for the formation of maximally
entangled Bell states of neutral atoms.

Phonon-mediated interactions between Rydberg atoms have been treated 
in a recently 
published paper by Gambetta \al~\cite{PhysRevLett.124.043402}. 
This work, however, focuses on multi-body interactions in optical 
lattices, while our paper focuses on phonon-mediated two-body 
interactions, and we demonstrate that these interactions can be 
made independent of the temperature of the atoms in direct analogy to 
the trapped ion case \cite{PhysRevLett.82.1971, kirchmair2009}.

Although the external degrees of freedom play a central role in the 
trapped ion quantum system, their use has not been fully explored 
in ultracold Rydberg platforms. The recent \cite{PhysRevLett.124.043402} 
work proposes the occurrence of non-binary interactions by 
electron-phonon coupling, while there has also been a number of works 
studying mediated interactions in self-assembled dipolar crystals 
\eg\cite{Buchler2007,PhysRevLett.100.050402}.
Here we aim for a scalable high fidelity platform for the creation 
of Bell states using trapped, neutral, Rydberg-dressed rubidium 
atoms for our qubits, and rely on the strong dipole-dipole 
interactions to induce motion, like in the M\o{}lmer-S\o{}rensen 
trapped ion gate, 
where entanglement is achieved via phonon mediated interactions 
\cite{PhysRevLett.92.207901,Friedenauer2008,PhysRevLett.103.120502,%
Kim2010,Richerme2014,Jurcevic2014,Zhang2017}.
This is realized by transient mapping of the qubit states of the atoms onto 
a mode of collective motion. At the end of the sequence, the 
qubit state is disentangled from the motion again 
\cite{PhysRevLett.82.1971,PhysRevLett.86.3907,PhysRevA.62.022311}.
This phonon-mediated interaction, treated in the original MS paper 
\cite{PhysRevLett.82.1971}, 
does not depend on the initial state of the phonon modes to lowest 
order. This makes for a reliable entanglement mechanism, even if the 
qubits are strongly coupled to a thermal reservoir 
\cite{PhysRevLett.82.1971}, and can possibly be used as the basis for 
a two-qubit quantum gate. 

Our approach starts with two (Rb) atoms with four distinct 
states each, two long lived states $|g_0\rangle$ and $|g_1\rangle$ 
and two Rydberg states $|r_0\rangle$ and $|r_1\rangle$, trapped in two 
well separated harmonic traps 
\begin{align}
V=\frac{1}{2}m\nu^2\left[(x_1-l/2)^2+(x_2+l/2)^2\right]
+V_\text{Ryd}(x_1-x_2),
\end{align}
with $x_j$ the position of atom $j$, $l$ the distance between the the 
oscillator minima, $m$ the mass of each atom and 
$V_\text{Ryd}(x)=C_6/x^6$ the state dependent, repulsive (in the case 
of rubidium $nS$-states) Rydberg-Rydberg van der Waals interaction.
This can be rewritten in relative and center-of-mass (CM) 
coordinates
\begin{align}
V=\frac{1}{2}\nu^2\left[m_r(r-l)^2+m_RR^2\right]
+V_\text{Ryd}(r), \label{eq:poteff}
\end{align}
where $m_r=m/2$ and $r$ are the reduced mass $m_1m_2/(m_1+m_2)$ 
and relative coordinates and $m_R=2m$ and $R$ are the CM mass and 
coordinate. 

\begin{figure}[h]
\centering
\includegraphics[width=.8\columnwidth]{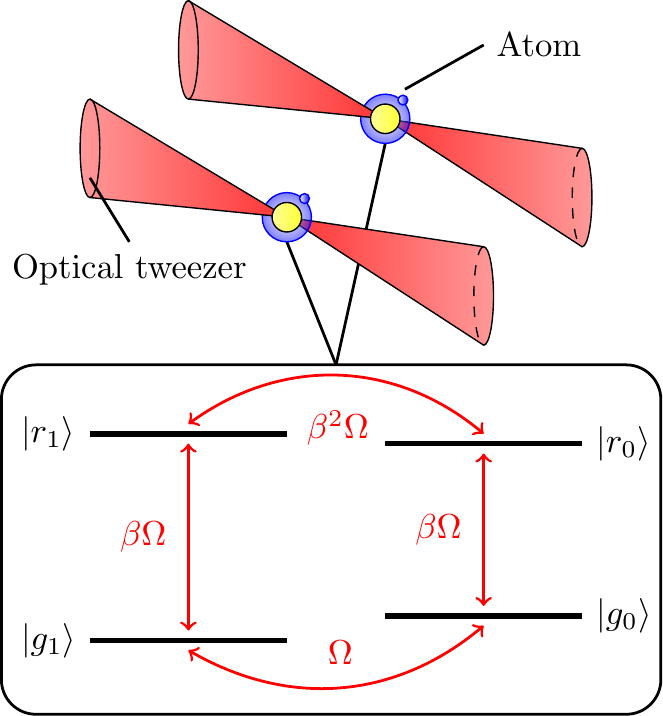}
\caption{Excitation scheme for using dressed qubits. The useful 
quantum states for qubits are the 'cross-dressed' states 
$|O\rangle = (|g_0\rangle+\beta|r_1\rangle)/\sqrt{1+\beta^2}$ and 
$|I\rangle = (|g_1\rangle+\beta|r_0\rangle)/\sqrt{1+\beta^2}$.
}
\label{fig:4lev}
\end{figure}

In order to lift the degeneracy of the CM and relative modes, we will use Rydberg 
dressed qubit states. 
We therefore apply the excitation scheme sketched in 
Fig.~\ref{fig:4lev}, where the four internal states of the proposed qubits are 
coupled via four lasers (which could effectively be a combination of lasers in case of a two-photon transition). The parameter $\beta$ is a small dimensionless 
number, indicating the ratio between the Rabi frequency for the ground state to ground state coupling connecting $|g_0\rangle$ and $|g_1\rangle$ and those of the 
dressing lasers, connecting $|g_j\rangle$ and $|r_j\rangle$ for 
$j=0,1$). Additionally, the Rabi frequency of the coupling laser 
that is connecting $|r_0\rangle$ and $|r_1\rangle$ is scaled by $\beta^2$. 
The interaction between the laser field and a single atom is described 
by the single atom Hamiltonian 
\begin{align}
\label{eq:sing_in}
{H}^{(1)}=&\frac{\hbar\Omega}{1+\beta^2}\times\\
\bigg(&e^{i(\eta_g(\hat{a}^\dagger+\hat{a})-\omega_gt)}|g_1\rangle\langle g_0|\nonumber\\
&+\beta e^{i(\eta_0(\hat{a}^\dagger+\hat{a})-\omega_0t)}|r_0\rangle\langle g_0|\nonumber\\
&+\beta e^{i(\eta_1(\hat{a}^\dagger+\hat{a})-\omega_1t)}|r_1\rangle\langle g_1|\nonumber\\
&+\beta^2 e^{i(\eta_{r}(\hat{a}^\dagger+\hat{a})-\omega_{r}t)}|r_1\rangle\langle r_0|
\bigg)\, +\,\textrm{H.C.},\nonumber
\end{align}
where $\beta$ is the dressing parameter, $\Omega$ is the Rabi 
frequency, $\eta_l=k_l\cdot \hat z\sqrt{\hbar/2m\nu}$ is the 
$l^\text{th}$ transition Lamb-Dicke parameter ($k_l$ is the wave number, 
$\hat z$ is a unit vector and $l=g,0,1,r$), $\hat{a}$ and 
$\hat{a}^\dagger$ are 
the ladder operators of the qubit trap 
and $\omega_l$ is the $l^\text{th}$ laser frequency. The exponential factors treat the 
effect of the lasers on the external/trap states, which we will initially 
ignore, and only consider their effect on the internal states, by expanding 
the exponentials in ${H}^{(1)}$ to zeroth order, denoted 
$\tilde{H}^{(1)}$. 

The zeroth order single atom Hamiltonian has two dark 
states 
\begin{align}
|D_0\rangle=&\frac{1}{\sqrt{1+\beta^2}}
\left(\beta|g_0\rangle-|r_0\rangle\right)\nonumber\\
|D_1\rangle=&\frac{1}{\sqrt{1+\beta^2}}
\left(\beta|g_1\rangle-|r_1\rangle\right),
\end{align}
which we will ignore, and two bright states 
\begin{align}
|O\rangle=&\frac{1}{\sqrt{1+\beta^2}}
\left(|g_0\rangle+\beta|r_1\rangle\right)\nonumber\\
|I\rangle=&\frac{1}{\sqrt{1+\beta^2}}
\left(|g_1\rangle+\beta|r_0\rangle\right),
\end{align}
which we will use as qubit states, as
$\tilde{H}^{(1)}|O\rangle=\hbar\Omega|I\rangle$ 
and $\tilde{H}^{(1)}|I\rangle=\hbar\Omega|O\rangle$.
Initialization of the qubit states can be performed by appropriate 
laser pulses.
Rydberg dressing gives longer life 
times of our qubit states, compared to direct Rydberg excitation, 
and allows for a finer tuning of the 
interaction strength by means of adjusting the dressing parameter 
$\beta$ in addition to choice of Rydberg state.

The interaction between the atoms and the laser light not only changes 
the internal state of the atom, but also their external state, 
\ie the atoms gain momentum. Therefore we have to consider the full laser 
interaction Hamiltonian Eq.~\eqref{eq:sing_in}, including the exponential 
factors. Using the shorthand notation 
$\theta_l=\eta_l(\hat{a}^\dagger+\hat{a})-\omega_lt$, and 
projecting ${H}^{(1)}$ onto the basis 
\begin{align}
S=(D_0,D_1,O,I)=\frac{1}{\sqrt{1+\beta^2}}
\begin{pmatrix}
0&-\beta&1&0\\
-\beta&0&0&1\\
1&0&0&\beta\\
0&1&\beta&0
\end{pmatrix},
\label{eq:basis}
\end{align}
constructed from the dark and qubit states of $\tilde{H}^{(1)}$, 
we get 
\begin{align}
S^{-1}{H}^{(1)}S&=\hbar\Omega\bigg(
\beta(e^{-i\theta_0}-e^{-i\theta_g})|D_0\rangle\langle O|\nonumber\\
&+\beta(e^{-i\theta_1}-e^{i\theta_g})|D_1\rangle\langle I|
+e^{i\theta_g}|O\rangle\langle I|
\bigg) + H.C.,
\label{eq:sing_in_rot}
\end{align}
ignoring terms higher than second order in $\beta$, since a realistic 
setup would be $nS$ Rydberg states with $n\approx 100$, 
$l\approx 3\,\mu$m and $\nu \approx 2\pi\times 100\,$kHz, we can 
expect $\beta<0.1$, as we will explain below.
Therefore neglecting these terms lead to errors on the 
order of 1\%.

Additionally, assuming $\eta_l$ to be small and 
taking the Lamb-Dicke approximation, we get 
\begin{align}
\Phi=&\exp\left[-i\eta_k (\hat{a}^\dagger+\hat{a})\right]
-\exp\left[-i\eta_l(\hat{a}^\dagger+\hat{a})\right]\nonumber\\
\approx&\,
i\left(\eta_l-\eta_k\right)(\hat{a}^\dagger+\hat{a})
-\frac{1}{2}\left(\eta_l^2-\eta_k^2\right)(\hat{a}^\dagger+\hat{a})^2.
\end{align}
We will here assume that $\eta_0$ and $\eta_1$ are not only small 
and comparable to $\eta_g\approx 0.05$, but 
in fact of equal absolute value. This is not only desirable, but also 
easily realizable as the Lamb-Dicke parameter can be tuned for two-photon 
transitions. With counter propagating dressing lasers and 
with Lamb-Dicke parameters close to that of the ground state to ground state coupling, 
$\eta_0=\eta_g+\xi=-\eta_1$ (with dimensionless $|\xi|<<\eta_g$), 
we can ensure 
that the exponential factors of the $|D_0\rangle\langle O|$ and 
$|D_1\rangle\langle I|$ terms in Eq.~\eqref{eq:sing_in_rot} are
limited in absolute value, to leading order, by
\begin{align}
\Phi\approx&\,
\beta\xi\left|i(\hat{a}^\dagger+\hat{a})-\eta_g(\hat{a}^\dagger+\hat{a})^2\right|
\nonumber\\
\lesssim&\, 
0.001\left|i(\hat{a}^\dagger+\hat{a}) - \eta_g(\hat{a}^\dagger+\hat{a})^2\right|,
\end{align}
which we can neglect, for reasonably low vibrational states 
\ie the CM mode quantum number $n_\textrm{R}<10$, as they 
contribute on the order of 1\% to the Hamiltonian, leaving
\begin{align}
S^{-1}&{H}^{(1)}S={\hbar\Omega}
e^{i(\eta_g(\hat{a}^\dagger+\hat{a})-\omega_gt)}|O\rangle\langle I|+H.C.
\label{eq:sing_in_red}
\end{align}
With these approximations, ${H}^{(1)}$ only cycles between the two-qubit 
states. Allowing for a detuning of the ground state to ground state coupling and one of the 
dressing lasers, we 
add $D=-\hbar\Delta(|g_1\rangle\langle g_1|+|r_0\rangle\langle r_0|)$ to 
$H^{(1)}$, resulting in the qubit detuning 
\begin{align}
S^{-1}DS=-\hbar\Delta(|D_0\rangle\langle D_0|+|I\rangle\langle I|).
\end{align}

This dressing makes the Van der Waals 
interaction between the two atoms independent of state, while having 
long life time compared to bare Rydberg atom qubits. The Van der 
Waals interactions will 
lift the degeneracy of the CM and relative modes of motion, as the 
oscillator frequency of the CM mode remains unchanged and the relative 
mode frequency increases. This results in a simplified Hamiltonian, 
in the absence of laser light,
\begin{align}
H_0=\hbar \bigg[\nu_r\left(a_r^\dagger \hat{a}_r +\frac{1}{2}\right)
&+\nu\left(\hat{a}_R^\dagger \hat{a}_R +\frac{1}{2}\right)\nonumber\\
&+\sum_{\sigma\in S} \omega_\sigma|\sigma\rangle\langle\sigma|
\bigg],
\end{align}
with $\nu_r$ the relative mode oscillator frequency, $\hat{a}_R$ ($\hat{a}_r$) and Hermitian 
conjugate are the CM (relative) mode ladder operators, the sum 
runs over the internal states and $\hbar\omega_\sigma$ is the energy of 
state $\sigma$.

\begin{figure}[ht]
\centering
\includegraphics[width=\columnwidth]{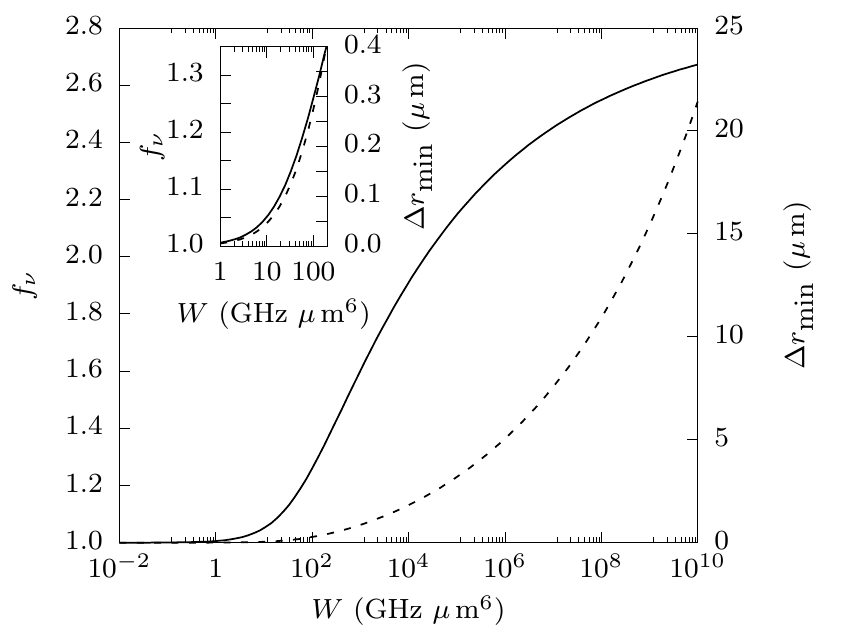}
\caption{
Ratio of relative and CM mode oscillator frequencies $f_\nu=\nu_r/\nu$
as function of the dressed interaction strength $W$ (solid) and the
corresponding shift in relative mode minimum position (dashed). Inset 
shows the range of small $W$ in greater detail.
\label{fig:nunu}}
\end{figure}

The inter-particle Rydberg-Rydberg interaction will only affect the 
relative mode. The relative frequency and the shift in 
the relative minimum position are also a function of the trapping 
frequency $\nu$ and the distance between the traps $l$, and 
to fully characterize the mode splitting we have to take all four 
parameters $\nu$, $l$, $C_6$, and $\beta$ into account. 
We introduce the dressed interaction strength 
\begin{align}
W=\beta^4C_6,
\end{align}
since the strength of the interaction between Rydberg dressed 
atoms is scaled by $\beta^4$ \cite{PhysRevA.95.043606}.

Ideally we would like to achieve a splitting ratio $\nu_r/\nu=\sqrt{3}$, 
as this would mimic to the ion-ion case. However, 
at the same time we have to minimize the shift in minimum 
position, realize sufficiently large life time (scaling with 
$\beta^{-2}$) and keep gate operation times low, 
therefore we have to consider splitting ratios smaller than $\sqrt{3}$. 
We find that the splitting needs to be larger than $1.15$, in 
order to make a reliable transfer with good fidelity.

For a given dressed interaction strength $W$ only one local minimum 
exists in the potential Eq.~\eqref{eq:poteff} for (real) 
positive relative coordinate, see Fig.~\ref{fig:nunu}. This 
minimum is located at $r_\textrm{min}$, which is the solution to 
\begin{align}
r_\textrm{min}^8-lr_\textrm{min}^7-6\frac{\hbar W}{\nu^2 m_r}=0.
\label{eq:rmin}
\end{align}
Expanding the potential around this minimum, we find the 
splitting ratio $f_\nu=\frac{\nu_r}{\nu}$ as 
\begin{align}
f_\nu=\sqrt{8-7\frac{l}{r_\textrm{min}}},\quad l>0,
\end{align}
which is shown in Fig.~\ref{fig:nunu}. 
Since $r_\textrm{min}$ grows monotonically for increasing $W$, the upper 
limit of the splitting fraction is $\sqrt{8}$ and lower limit is 1. 
This gives us a large range of controllable splitting fraction,  
limited the distance between the single atom traps.
Inversely, it is more convenient to determine 
what strength is needed to result in a sufficient splitting fraction 
and the shift in minimum position can then be determined as
\begin{align}
r_\textrm{min}=\frac{7l}{8-f_\nu^2},\quad l>0,
\end{align}
from which $W$ can be derived, using Eq.~\eqref{eq:rmin}. 
This treatment is limited by the validity of the harmonic approximation 
of the effective potential around the local minimum $r_\textrm{min}$. 
However, for reasonable values of $l$ and $W$ the approximation holds 
for a large range around the minimum and a large number of bound states 
are consistent with this approximation.

\begin{figure*}[t]
\includegraphics[width=0.95\textwidth]{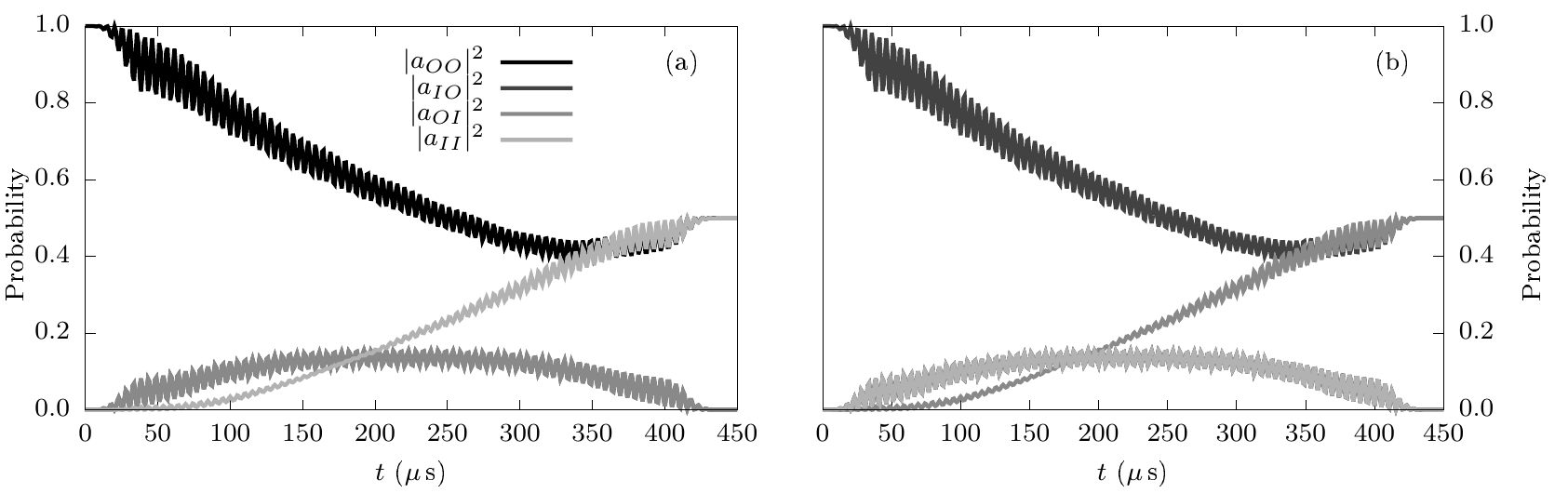}
\includegraphics[width=0.95\textwidth]{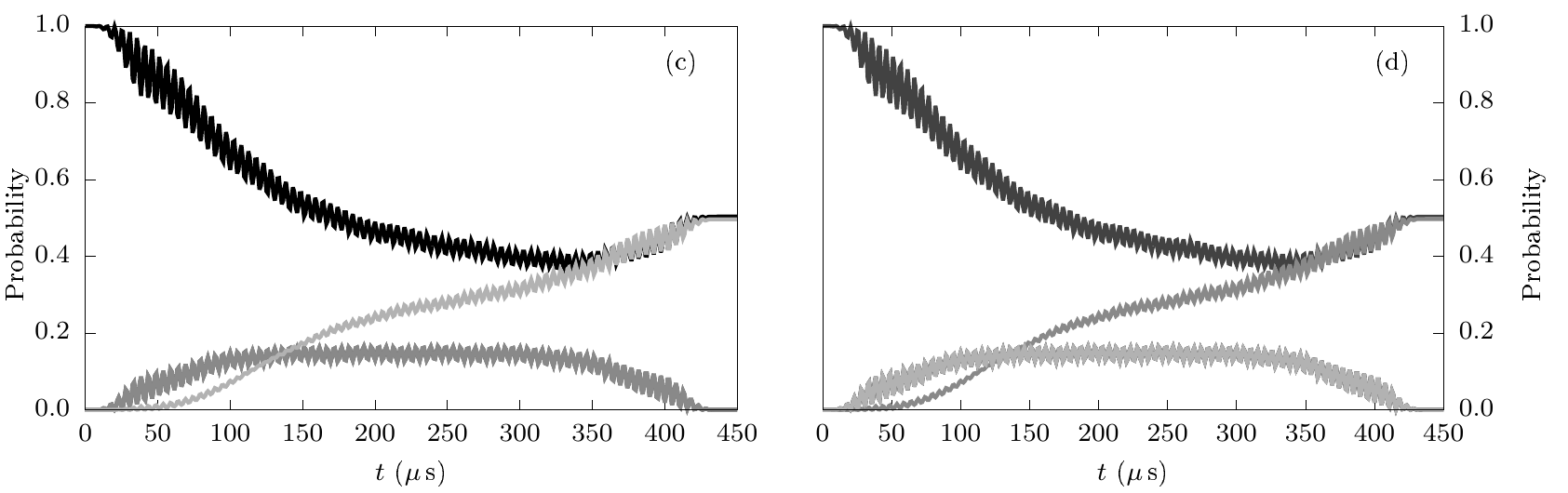}
\caption{ \label{fig:bell}
Time evolution at finite temperature of the two-qubit populations with 
the vibrational motion traced out, obtained by propagating $H_I$ 
(Eq.~\eqref{eq:Hint}). Labels denote the population of the 
indicated state, $OO$ is the population of $|OO\rangle$ and so forth. 
The subfigures show 
populations of the time evolved states resulting in the creation of 
Bell states from pure qubit states 
$|OO\rangle\rightarrow \frac{1}{\sqrt{2}}(|OO\rangle+i|II\rangle)$ 
(a and c) and 
$|IO\rangle\rightarrow \frac{1}{\sqrt{2}}(|IO\rangle+i|OI\rangle)$ 
(b and d) at 0 temperature (a and b) and $5\,\mu$K 
(c and d). The other input states mirror this behavior. These Bell 
states can be achieved with high fidelity at realistic cold atom 
temperatures, given the approximations made in this paper. 
}
\end{figure*}
We induce spin-spin interactions by letting both qubits interact 
with bichromatic laser light, slightly detuned both above and below 
resonance. 
Including the photon recoil in the dressed frame, 
the effect of the laser light acting on both qubits, each with internal 
states $|O\rangle$ and $|I\rangle$ and trap states 
$|n_R,n_r\rangle$, where $n_R$ ($n_r$) is the CM (relative) mode 
vibrational quantum number, can be expressed in the two-qubit 
Hamiltonian 
\begin{align}
H^{(2)}&=\sum_j\sum_{k_j}\frac{\hbar\Omega_{k_j}}{2}
\sigma_{+,j}e^{-i\omega_{k_j}t}\times\nonumber\\
&\exp\left[i\eta_{k_j}\left(\hat{a}_R^\dagger+ \hat{a}_R
-\frac{(-1)^j\sqrt{f_\nu}}{2}(a_r^\dagger+ \hat{a}_r)\right)
\right]\nonumber\\
&+H.C.,
\end{align}
where $k_j$ is used to label the laser beams interacting with the 
$j$ atom and $\sigma_{\pm,j}$ are the internal state step operator 
for atom $j$. We will use a sufficiently large mode splitting such 
that the relative mode is effectively frozen out. 
Changing to the interaction picture, we define the rotated 
creation operator 
\begin{align}
\hat{b}_j=e^{-i\nu t}\hat{a}_R 
-\frac{(-1)^j\sqrt{f_\nu}}{2}e^{-i\nu_r t}\hat{a}_r,
\label{eq:boperator}
\end{align}
where $j=1,2$ is the atom site number, and get the two-qubit 
interaction picture Hamiltonian 
\begin{align}
H_\text{I}&=\frac{\hbar}{2}\sum_j 
\sum_{k_j}\Omega_{k_j}e^{i\delta_{k_j}t}\sigma_{+,j}\nonumber
e^{i\eta_{k_j}\left( \hat{b}^\dagger_j+ \hat{b}_j\right)}\\
&+\Omega_{k_j}e^{-i\delta_{k_j}t}\sigma_{-,j}
e^{-i\eta_{k_j}\left( \hat{b}^\dagger_j+ \hat{b}_j\right)},
\label{eq:Hint}
\end{align}
where $\delta_{k_j}=\omega_{k_j}-\omega_{O\rightarrow I}$ is the 
detuning from the $|O\rangle\rightarrow |I\rangle$ transition. 
This Hamiltonian Eq.~\eqref{eq:Hint} reduces to a spin-spin 
interaction Hamiltonian 
\cite{PhysRevLett.103.120502,PhysRevLett.92.207901,Roos_2008}.
Assuming $\eta_{k_j}=\eta$,
$\Omega_{k_j}=\Omega$ 
and $|\delta_{k_j}|=\delta\approx \nu$, we can, in the Lamb-Dicke 
limit, simplify the interaction picture Hamiltonian 
\begin{align}
H_\text{I}&\approx 
\frac{\hbar\Omega}{2}\sum_j \sum_{k_j}
e^{i\delta_{k_j}t}\sigma_{+,j}\nonumber
\left[1+i\eta\left( \hat{b}^\dagger_j+ \hat{b}_j\right)\right]+H.C.,
\end{align}
which we further simplify by using
\begin{align}
e^{i\delta_{k_j}t} \hat{b}_j=e^{i(\delta_{k_j}-\nu)t}\hat{a}_R
-\frac{(-1)^j\sqrt{f_\nu}}{2}e^{i(\delta_{k_j}-\nu_r) t} \hat{a}_r,
\end{align}
and by neglecting fast rotating terms. 
This results in
\begin{align}
H_\text{I}\approx&\Omega\bigg(2\hbar\cos(\delta t)J_x
-\sqrt{2\hbar\nu m_R}\eta J_y\bigg(\cos(\nu t-\delta t)R\nonumber\\
&+\frac{\sin(\nu t-\delta t)}{m_R\nu}p_R
\bigg)
\bigg)-H_r,
\label{eq:redHam}
\end{align}
where $J_x$ and $J_y$ are collective spin operators, $p_R$ is the 
CM mode momentum operator and
\begin{align}
H_r=&\sqrt{\frac{\hbar m_r \nu_r}{2}}\Omega\eta_r
\tilde{ J_y}
\bigg(\cos(\nu_r t-\delta t)r\nonumber\\
&+\frac{\sin(\nu_r t-\delta t)}{m_r\nu_r}p_r\bigg),
\end{align}
with $p_r$ the relative mode momentum operator, 
$\tilde J_y=\sigma_{y,2}-\sigma_{y,1}$ 
and $\eta_r=\eta/\sqrt{f_\nu}$ the relative mode Lamb-Dicke parameter. 
If we ignore the fast rotating $J_x$ term, we can write the 
propagator by virtue of the Zassenhaus formula 
\begin{align}
U(t)&=\exp\left[-i \frac{\eta^2\Omega^2}{\nu-\delta} J_y^2 A(t)\right]\,
\exp\left[-i \frac{\eta_r^2\Omega^2}{\nu_r-\delta} \tilde J_y^2 B(t)\right]\nonumber\\
&\times
\exp\left[-i\alpha_R J_y \frac{\sin (\nu-\delta)t}{\nu-\delta}R\right]\,
\nonumber\\
&\times
\exp\left[-i\alpha_R J_y \frac{1-\cos (\nu-\delta)t}{m_R\nu(\nu-\delta)}p_R\right]
\nonumber\\
&\times
\exp\left[i\alpha_r \tilde J_y \frac{\sin (\nu_r-\delta)t}{\nu_r-\delta}r\right]\,
\nonumber\\
&\times
\exp\left[i\alpha_r \tilde J_y 
\frac{1-\cos (\nu_r-\delta)t}{m_r\nu_r(\nu_r-\delta)}p_r\right],
\label{eq:propagator}
\end{align}
with $\alpha_R=\sqrt{\frac{2m_R\nu}{\hbar}}\eta\Omega$ and 
$\alpha_r=\sqrt{\frac{2m_r\nu_r}{\hbar}}\eta_r\Omega$. $A(t)$ and $B(t)$ 
can be determined from the Schr\"odinger equation similar to $A(t)$ in 
Eq.~(9) of ref.~\cite{PhysRevA.62.022311}.

At times $\tau_k=2k\pi/(\nu-\delta)$ (with $k$ an integer), the 
propagator Eq.~\eqref{eq:propagator} reduces to that of 
a spin-spin Hamiltonian 
\begin{align}
U&(\tau_k)\approx
\exp\left({-i \frac{\eta^2\Omega^2 }{\nu-\delta}J_y^2 A(\tau_k)}\,
{-i \frac{\eta_r^2\Omega^2}{\nu_r-\delta}\tilde J_y^2 B(\tau_k)}\right),
\label{eq:red_propa}
\end{align}
with exact equality if $\nu(f_\nu-1)/(\nu-\delta)$ is an integer, 
however, the approximation always has merit if  
\mbox{$\frac{\nu_r-\delta}{\nu-\delta}>>1$}. 

We apply the Hamiltonian Eq.~\eqref{eq:Hint} in the time-dependent 
Schr\"odinger equation in the interaction picture 
with interaction picture state $\psi_\text{I}$ \cite{Porras04,Kim09}. We set the 
Rabi frequency such that 
\begin{align}
\Omega=\frac{\nu-|\delta|}{2\eta}\cdot (1+\alpha)\cdot\mathcal{T}(t),
\label{eq:ratio}
\end{align}
with alpha being a small dimensionless number and 
\begin{align}
\mathcal{T}(t)=\begin{cases}
\sin^2\left(\frac{\pi t}{2 t_\textrm{s}}\right) & t<t_\textrm{s}\\
1&t_\textrm{s}<t<t_\textrm{p}-t_\textrm{s}\\	
\cos^2\left(\pi\frac{ t+t_\textrm{s}-t_\textrm{p}}{2 t_\textrm{s}}\right)
&t_\textrm{p}-t_\textrm{s}<t<t_\textrm{p}\\
0&\textrm{otherwise}
\end{cases}
\end{align} 
is a ramping function with $t_\textrm{s}$ being the ramping time and 
$t_\textrm{p}$ is the length of the pulse. 
In the original MS paper \cite{PhysRevLett.82.1971} 
$\alpha$ is zero, as this ensures a $\pi/2$ rotation in 
phase space, but small adjustments to the Rabi frequency must be made to 
compensate for the (usually) weaker mode splitting achieved with the 
Rydberg interaction. 

We have simulated the coherent time evolution starting from each 
of the four two-qubit states 
($|OO\rangle$, $|OI\rangle$, $|IO\rangle$ and $|II\rangle$), in 
combination with a thermal ensemble of oscillator states at  
temperatures ranging from $0\,\mu$K to $5\,\mu$K, see 
Fig.~\ref{fig:bell} for examples.  
For this simulation, we have set all Lamb-Dicke parameters to 
$\eta=0.05$, the detunings are set to $\delta=\pm 0.975\nu$,  
the dressed interaction strength $W=50$ GHz$\,\mu$m$^6$, the trap 
frequency is $\nu=2\pi\times100\,$kHz and the distance between the 
atoms is set to $l=3\,\mu$m. 
The resulting splitting fraction is $f_\nu=1.1745$ and atoms are 
pushed a further $r_\textrm{min}=0.1719\,\mu$m apart. 
In order to account for 
the off-resonant phase accumulation in the relative mode of motion, which 
is much closer in frequency compared to the trapped ion case, we
need $\alpha=0.1333$.

Our simulation shows reliable creation of Bell states, at all 
temperatures starting from all four of the internal two-qubit states.
Tracing out the vibrational states, we find fidelties of Bell state 
creation
to be higher than $0.999$ for all input states even at 
non-zero temperature, under the approximations given above. 
We expect both the anharmonicity of the trap and 
non-magic trapping of the Rydberg part \cite{PhysRevA.97.013430} 
to influence the 
fidelity of the entanglement mechanism negatively: We estimate the 
trap quality issues to reduce the fidelity of Bell state creation by 
$\sim2\%$. Further we expect the finite life-time 
of the Rydberg-dressed qubits, which we estimate to influence 
the overall fidelity by $\sim1\%$ for the $100S$ Rydberg 
level in rubidium-85. Additional losses and reductions in 
fidelity, due to neglected terms in the Hamiltonian are all below 
1\%, as they are all higher order in $\beta\sim 0.1$ or 
$\eta\sim0.01$. By increasing the principal quantum number of the 
Rydberg level $n$ of the dressed 
qubits, we expect these approximations to have a smaller effect on 
the overall fidelities, as $\beta\propto n^{-11/4}$. 
The lifetime of the Rydberg-dressed state will also increase 
\cite{He1990,Branden2009,PhysRevA.79.052504,PhysRevA.95.043606} 
as $\beta^{-2}n^3\propto n^{8}\sqrt{n}$ by neglecting black body 
radiation, which of course limits the lifetime, but is not 
detrimental to this analysis, and can be reduced by means of 
a cryostat. This leaves only the quality of the traps as a significant 
source of errors, which can not simply be reduced by 
a change of the dressing parameter, and we expect this will 
be the limiting factor. 

Recent years have seen many implementations of 
single atom traps, like optical tweezers  
\cite{Labuhn2016,Barredo2016,PhysRevX.2.041014}, 
holographic trapping \cite{Xu2010,PhysRevX.4.021034}, photonic 
crystals trapping \cite{Yu2014}, cavity trapping 
\cite{Pinkse2000,PhysRevLett.83.4987}, magneto optical traps 
\cite{Yoon2007}
or magnetic microtraps \cite{WANG20161097,PhysRevA.97.013430}. 
Both magnetic microtraps and optical tweezer arrays 
\cite{Labuhn2016,Barredo2016} can be very tight with frequencies in 
the $10-100$kHz and the separation of two trap sites is on the 
$\mu$m scale. 
This development of tight single-atom traps with high filling factor 
forms the main motivation of this paper to investigate the MS gate 
for dressed Rydberg atoms.
An interesting future development would be to employ a trapped 
ion crystal to mediate interactions between atomic qubits. 
This would combine 
long-range Coulomb interactions with the favorable scaling properties 
of neutral quantum devices \cite{PhysRevA.94.013420}.

In this paper, we have shown that it should be possible to implement a 
M\o{}lmer-S\o{}rensen gate between two atoms trapped in tweezers. 
Combined with single qubit gates, the MS gate forms a universal set of 
quantum gates that has been implemented in trapped ions with very high 
fidelity 
\cite{PhysRevLett.82.1971,PhysRevLett.86.3907,PhysRevA.62.022311}. 
Our work shows, that it should be possible to extend its use to neutral 
atomic systems, that have much better scalability prospects. 
We have shown that, by appropriate choices of Rydberg level and dressing 
parameters, it is possible to create maximally entangled states 
with qubits consisting of Rydberg-dressed atoms in a 
Boltzmann-distributed statistical mixture of oscillator states, with 
experimentally realistic laser parameters, 
and we have quantified the order of magnitude of 
the errors.
Besides the quantum gate described in this work, the scheme may be 
beneficial for the creation of atomic quantum simulators of quantum spin 
models \cite{Weimer2010}. 
Here the tweezer setup offers in particular the benefit of creating 
nearly arbitrary trapping geometries \cite{PhysRevLett.124.043402}.

During the preparation of this paper, we became aware of a related 
work by Gambetta \al\cite{PhysRevLett.124.043402}, which focuses on 
many-body interactions in tweezer arrays. Our work has been conducted 
independently of Gambetta \al and focuses instead on two-body 
interactions.


\begin{acknowledgments}
This research was financially supported by the Foundation for 
Fundamental Research on Matter (FOM), and by the Netherlands 
Organization for Scientific Research (NWO). We also acknowledge 
the European Union H2020 FET Proactive project RySQ (grant N. 
640378). RG and SK acknowledge support by Netherlands Organization for 
Scientific Research (Vrije Programma 680.92.18.05). RG acknowledges support 
by the Netherlands Organization for Scientific 
Research (Vidi Grant 680-47-538 and Start-up grant 740.018.008).
\end{acknowledgments}

\bibliography{main}
\clearpage
\end{document}